\documentclass[amsmath,11pt]{article}

\usepackage{amssymb}

\date{\empty}

\begin{document}

\title{\bf The peculiar Raychaudhuri equation}

\author{Christos G. Tsagas and Miltiadis I. Kadiltzoglou\\ {\small Section of Astrophysics, Astronomy and Mechanics, Department of Physics}\\ {\small Aristotle University of Thessaloniki, Thessaloniki 54124, Greece}}

\maketitle

\begin{abstract}
Peculiar motions are commonplace in the universe. Our local group of galaxies, for example, drifts relative to the Hubble flow at about 600~km/sec. Such bulk flows are believed to fade away as we move on to progressively larger scales. Recently, however, there have been reports of peculiar motions larger and faster than typically expected. If these claims are correct, the role of peculiar flows in shaping the kinematics of our universe has been probably underestimated. Here, we use general relativistic techniques to analyse the average kinematics of large-scale bulk motions and compute the nonlinear Raychaudhuri equation of such flows. In doing so, we introduce two families of observers. One at rest with the Hubble expansion and another following the peculiar motion. We first derive the fully nonlinear expressions of the ``peculiar'' Raychaudhuri equation in both frames and identify a range of relative motion effects. Linearised around a Friedmann universe with pressureless dust, the full equations reduce to a simple relation with no explicit contribution from the matter component of the universe. Using cosmological perturbation theory, we obtain a scale-dependent formula, which also incorporates the effects of universe's inhomogeneity and anisotropy. When the latter are secondary, the peculiar Raychaudhuri equation can be solved analytically. Also, combining the resulting expression with data from recent peculiar-velocity surveys, we estimate the relative deceleration/acceleration rates of such bulk motions.
\end{abstract}

\section{Introduction}\label{sI}
Realistic observers do not simply follow the mean expansion of the universe, but have their own peculiar velocities. The Milky Way and the nearby galaxies, for example, ``drift'' relative to the Hubble flow with a speed that exceeds 600~km/sec. Our Local Group is not an exception, since large-scale bulk motions appear to be the norm. In fact, during the last five years, there have been claims of bulk flows much faster and much larger than generally expected~\cite{WFH}-\cite{AF}. The surveys of~\cite{WFH}, in particular, report peculiar motions around 400~km/sec on scales of the order of 100~Mpc. The studies of~\cite{KA-BKE,AF} cover much larger areas (of Gpc-order) and report speeds that approach, or even surpass, the 1000~km/sec threshold. The recently released Planck data (see~\cite{Aetal1} for further discussion) seem to exclude large-scale peculiar motions as fast as those reported in~\cite{KA-BKE,AF}, see however~\cite{A-B}, but they stop short from constraining those of~\cite{WFH}.

Relative motion effects can change the observers' perception of what we may call ``reality'' and can lead to apparent paradoxes in some cases. For instance, peculiar motions introduce a preferred direction in the 3-dimensional space that could be misinterpreted as a sign of spatial anisotropy by the unsuspecting observer. The dipole seen in the spectrum of the Cosmic Microwave Background (CMB) is probably the best known example of such an apparent anisotropy, triggered by our motion relative to the mean Hubble flow. Similar, though much weaker, dipole-like anisotropies have been recently reported to reside in a host of different cosmological data~\cite{SW}. Intriguingly, these dipoles appear closely aligned with their CMB counterpart. These are early days, however, and the data is still inconclusive. Nevertheless, it is conceivable that the reported close alignment of the dipoles may not be a mere coincidence, but the sign of a common origin. It might therefore prove useful to look further into the role and the implications of such large-scale bulk motions, as they might interfere with the observations and their interpretation~\cite{T}. Here, we take a step in this direction by considering the average peculiar kinematics and more specifically the ``peculiar'' Raychaudhuri equation.

Changes in the mean separation between a family of curves, namely whether they approach or move apart from each other, are monitored by Raychaudhuri's formula. In most cases, these curves are the timelike worldlines of observers living in the pseudo-Riemannian spacetimes adopted by General Relativity. It is also possible, however, to consider a congruence of spacelike curves and investigate whether these expand or contract and whether their expansion/contrcation speeds up or slows down. The latter requires deriving the corresponding Raychaudhuri equation. In our case, the aforementioned spacelike curves are tangent to the velocity field of the observers' peculiar flow relative to the universal expansion.
We describe this ``drift'' motion by adopting two coordinate systems: one following the mean expansion of the universe and another moving along with the peculiar flow. This makes it easier to identify the relative motion effects. Our equations, which are initially fully nonlinear, show that the matter content of the universe has little or no direct effect on the average peculiar kinematics. More specifically, the contribution of the local gravitational field to the ``peculiar version'' of Raychaudhuri's formula either changes nature or disappears altogether. This marks a considerable qualitative difference relative to the ``standard'' Raychaudhuri equation, where the Ricci-curvature plays the decisive role.

Assuming non-relativistic peculiar velocities, we linearise our equations around a (decelerating) Friedmann-Robertson-Walker (FRW) universe that contains nothing else but conventional pressureless dust. The linearisation takes place in the frame of the peculiar motion, which defines the coordinate system of any realistic observer in our universe. At the linear perturbative level, all the relative motion effects are encoded in  the ``peculiar acceleration'', which is primarily associated to the model's inhomogeneity and anisotropy. Turning to cosmological perturbation theory, we recast the peculiar Raychaudhuri equation in a scale-dependent form, which also includes (linear) density perturbations and shear distortions. The nature of our final formula, especially its scale-dependence, allows for a variety of possible results that may differ quantitatively as well as qualitatively. It appears that the main players are the background expansion and the universe's spatial inhomogeneity/anisotropy. The former typically acts as a source of cosmic ``friction'', which increases on progressively smaller lengths, but the role of latter is less straightforward to decode. On scales where the non-uniformity of the universe is negligible and the background expansion dominates, the peculiar flow slows down. This is not necessarily the case when the inhomogeneities and the anisotropies are also accounted for. Hence, we can in principle have bulk flows that expand (or contract) at a decelerating (or accelerating) pace.

When only the friction-like effects of the background expansion are included, the peculiar Raychaudhuri equation can be solved analytically. The solution enables us to quantify the decelerating effect of the Hubble flow on the mean peculiar kinematcs, as well as its scale dependence. Also, combining our theoretical formulae with data from recent surveys of large-scale peculiar velocities, we can estimate the deceleration rate of these bulk flows (relative to that of the background universe), as measured by observers living inside these drifting domains. Recall that our unperturbed model is a decelerating FRW cosmology filled with pressure-free dust. We find that drift motions with a size of $\sim100$~Mpc generally appear to slow down faster than the Hubble flow. Further out, on Gpc-order scales, the relative-deceleration ratio seems to drop by two orders of magnitude. Bulk motions that are large enough to encompass a few hundred Mpc and decelerate as fast as the background universe (or even faster) are an intriguing theoretical possibility, because (unsuspecting) observers living inside such regions are more likely to misinterpret their observational data than those moving along with the smooth Hubble flow.

\section{The Raychaudhuri equation}\label{sRE}
The Raychaudhuri equation is an essentially geometrical relation, monitoring changes in the mean separation between a congruence of curves that are tangent to a given vector field. In General Relativity, this congruence usually corresponds to a family of timelike worldlines and Raychaudhuri's formula monitors variations in their mean expansion or contraction.

\subsection{Fundamental observers}\label{ssFOs}
Let us consider a general 4-dimensional spacetime and introduce a family of fundamental observers, moving along the timelike 4-velocity field $u_a$. Assuming that $g_{ab}$ is the spacetime metric, the symmetric spacelike tensor $h_{ab}=g_{ab}+u_au_b$ projects orthogonal to $u_a$ (i.e.~$h_{ab}u^b=0$) and into the observers' instantaneous 3-D rest-space. The projector also defines the covariant derivative, ${\rm D}_a=h_a{}^b\nabla_b$, operating in the aforementioned rest-space, with $\nabla_a$ representing its 4-D counterpart. Then, the kinematics of the fundamental observes are described by four irreducible variables, obtained after decomposing the gradient of the $u_a$-field as follows (e.g.~see~\cite{TCM})
\begin{equation}
\nabla_bu_a= {1\over3}\,\Theta h_{ab}+ \sigma_{ab}+ \omega_{ab}- A_au_b\,.  \label{Nbua}
\end{equation}
In the above, $\Theta=\nabla^au_a={\rm D}^au_a$ is the volume scalar, $\sigma_{ab}={\rm D}_{\langle b}u_{a\rangle}$ is the shear tensor, $\omega_{ab}={\rm D}_{[b}u_{a]}$ is the vorticity tensor and $A_a=u^b\nabla_bu_a$ is the 4-acceleration vector.\footnote{Round brackets denote symmetrisation, square antisymmetrisation and angled brackets indicate the symmetric and trace-free component of second-rank tensors (e.g.~$\sigma_{ab}={\rm D}_{\langle b} u_{a\rangle}={\rm D}_{(b}u_{a)}- ({\rm D}^cu_c/3)h_{ab}$).} These are the irreducible kinematic variables of the $u_a$-field. The volume scalar describes the average separation between the worldlines of the $u_a$-congruence, namely the average expansion or contraction of the associated observers. The shear represents kinematic anisotropies, the vorticity monitors the rotational behaviour of the $u_a$-field and the 4-acceleration reflects the presence of non-gravitational forces. This means that $A_a$ vanishes when dealing with geodesic worldlines.

\subsection{The ``standard'' Raychaudhuri equation}\label{ssSRE}
The Raychaudhuri equation describes the (proper) time evolution of the volume scalar ($\Theta$) and follows by applying the Ricci identity to the fundamental 4-velocity field. Thus, our starting point is the relation
\begin{equation}
2\nabla_{[a}\nabla_{b]}u_c= R_{abcd}u^d\,,  \label{Ricci1}
\end{equation}
with $R_{abcd}$ representing the Riemann curvature tensor. Contracting the above along $u_a$ and then taking the trace of the resulting expression leads to Raychaudhuri's formula
\begin{equation}
\dot{\Theta}= -{1\over3}\,\Theta^2- \sigma_{ab}\sigma^{ab}+ \omega_{ab}\omega^{ab}+ {\rm D}^aA_a+ A^aA_a- R_{ab}u^au^b\,. \label{Ray1}
\end{equation}
Here, overdots indicate proper-time derivatives in the $u_a$-frame (i.e.~$\dot{\Theta}=u^a\nabla_a\Theta$) and $R_{ab}=R^c{}_{acb}$ is the Ricci tensor that carries the effects of the local gravitational field. Recalling that $R_{ab}u^au^b=(\rho+3p)/2-\Lambda$, expression (\ref{Ray1}) assumes the more familiar form
\begin{equation}
\dot{\Theta}= -{1\over3}\,\Theta^2- {1\over2}\,(\rho+3p)- \sigma_{ab}\sigma^{ab}+ \omega_{ab}\omega^{ab}+ {\rm D}^aA_a+ A^aA_a+ \Lambda\,, \label{Ray2}
\end{equation}
where $\rho$ is the density of the matter, $p$ is the (isotropic) pressure and $\Lambda$ is the cosmological constant.\footnote{Throughout this report we use geometrised units, with $\kappa=8\pi G=1=c$.} Note that the sum $(\rho+3p)/2$ provides the effective energy density of the matter fields and takes positive values for all known types of media.

Raychaudhuri's formula describes the mean expansion (or contraction) of a self-gravitating medium and has played a central role in the formulation of the various singularity theorems, as well as in modern cosmology. Following (\ref{Ray2}), negative terms on the right-hand side accelerate the contraction of a collapsing fluid, or slow down its expansion. Positive terms, on the other hand, do the opposite. Thus, applying Eq.~(\ref{Ray2}) to a converging family of irrotational timelike geodesics, one finds that the associated worldlines intersect within finite time. This leads to caustic formation and subsequently to a singularity~\cite{HE}. When applied to cosmology, on the other hand, the Raychaudhuri equation ensures that conventional matter (with $\rho+3p>0$) always decelerates the expansion. This result has been the main reason for introducing dynamical dark energy (or a positive cosmological constant) in an attempt to explain the supernovae data and the inferred accelerated expansion of the universe~\cite{Retal}.

In the literature one can find studies investigating the form of Raychadhuri's formula in a variety of physical environments and discussing its potential implications (see~\cite{CG} for a representative though incomplete list). To the best of our knowledge, all the available studies consider a single family of observers/frames and none allows for peculiar motions relative to the fundamental frame. In what follows, we will derive and discuss the Raychaudhuri equation associated with the peculiar motion of an observer family, drifting relative to a given reference frame. We will begin by considering a general spacetime with no symmetries. Once the nonlinear expressions have been obtained, we will turn to cosmology and linearise our formulae around an FRW universe with pressureless dust. Our aim is to study the average peculiar kinematics of observers in a typical galaxy, like the Milky Way for example, drifting relative to the smooth Hubble flow.

\section{The peculiar kinematics}\label{sPKs}
In the previous section we looked at a family of timelike worldlines and considered changes in their average separation (i.e.~expansion/contraction). This lead to expression (\ref{Ray2}) and to the general conclusions derived from it. One can apply the same process to a spacelike congruence as well. So, next, we will consider observers moving with respect to a reference (fundamental) 4-velocity field and obtain the Raychaudhuri equation of their ``peculiar'' motion.

\subsection{Drifting observers}\label{ssDOs}
In cosmology there is a preferred coordinate system, with respect to which we can define and measure peculiar velocities. This is the frame of the smooth Hubble expansion, which is by definition the coordinate system where the CMB dipole vanishes. Let us introduce a family of drifting observers, with peculiar velocity $v_a$, relative to such a reference 4-velocity field $u_a$. Their overall motion is described by the familiar Lorentz boost
\begin{equation}
\tilde{u}_a= \gamma(u_a+v_a)\,,  \label{Lorentz}
\end{equation}
with $\gamma=1/\sqrt{1-v^2}$ being the Lorentz-boost factor~\cite{KE}. In addition, $u_av^a=0$ by construction and $v^2=v_av^a$ by definition. The $u_a$ and $\tilde{u}_a$ fields are both timelike, namely $u_au^a=-1=\tilde{u}_a\tilde{u}^a$, and define the time direction of the corresponding observers. To distinguish between the two, we will use overdots to denote time differentiation in the (reference) $u_a$-frame (i.e.~~${}^{\cdot}=u^a\nabla_a$) and primes for time derivatives in the tilded frame (i.e.~${}^{\prime}= \tilde{u}^a\nabla_a$). At the same time, the tensors $h_{ab}=g_{ab}+u_au_b$ and $\tilde{h}_{ab}=g_{ab}+ \tilde{u}_a\tilde{u}_b$ project orthogonal to $u_a$ and $\tilde{u}_a$ respectively and into the instantaneous 3-D rest-spaces of the associated observers. Therefore, ${\rm D}_a=h_a{}^b\nabla_b$ and $\tilde{\rm D}_a=\tilde{h}_a{}^b\nabla_b$ define the corresponding 3-D covariant derivative operators. Finally, we monitor the kinematics of the $\tilde{u}_a$-field by means of the irreducible decomposition
\begin{equation}
\nabla_b\tilde{u}_a= {1\over3}\,\tilde{\Theta}\tilde{h}_{ab}+ \tilde{\sigma}_{ab}+ \tilde{\omega}_{ab}- \tilde{A}_a\tilde{u}_b\,,  \label{tNbtua}
\end{equation}
where $\tilde{\Theta}= \nabla^a\tilde{u}_a=\tilde{\rm D}^a\tilde{u}_a$, $\tilde{\sigma}_{ab}=\tilde{{\rm D}}_{\langle b} \tilde{u}_{a\rangle}$, $\tilde{\omega}_{ab}=\tilde{{\rm D}}_{[b}\tilde{u}_{a]}$ and $\tilde{A}_a= \tilde{u}^b\nabla_b\tilde{u}_a$. These are respectively the volume scalar, the shear tensor, the vorticity tensor and the 4-acceleration vector in the tilded (the drifting) frame.

\subsection{The drift flow}\label{ssDF}
Earlier, in \S~\ref{ssFOs}, we introduced a decomposition of the fundamental 4-velocity field, which led to the associated irreducible kinematic variables. The same splitting was applied to the 4-velocity of the tilded observers in \S~\ref{ssDOs} above. When studying relative motions, it helps to apply an analogous decomposition to the projected gradient of the (spacelike) $v_a$-field and thus obtain the kinematic components of the drift flow. As before, one should first specify the reference frame. Splitting the projected velocity gradient ${\rm D}_bv_a=h_a{}^ch_b{}^d\nabla_dv_c$, leads to
\begin{equation}
{\rm D}_bv_a= {1\over3}\,\vartheta\,h_{ab}+ \varsigma_{ab}+ \varpi_{ab}\,,  \label{Dbva}
\end{equation}
with $\vartheta={\rm D}^av_a$, $\varsigma_{ab}= {\rm D}_{\langle b}v_{a\rangle}$ and $\varpi_{ab}= {\rm D}_{[b}v_{a]}$~\cite{ET}. These are the irreducible kinematic variables of the peculiar motion, as measured in the reference $u_a$-frame. Alternatively, we may consider the gradient $\tilde{\rm D}_bv_a=\tilde{h}_a{}^c\tilde{h}_b{}^d \nabla_dv_c$ and introduce the decomposition
\begin{equation}
\tilde{\rm D}_bv_a= {1\over3}\,\tilde{\vartheta}\tilde{h}_{ab}+ \tilde{\varsigma}_{ab}+ \tilde{\varpi}_{ab}\,.  \label{tDbva}
\end{equation}
This time, $\tilde{\vartheta}=\tilde{\rm D}^av_a$, $\tilde{\varsigma}_{ab}= \tilde{\rm D}_{\langle b}v_{a\rangle}$ and $\tilde{\varpi}_{ab}= \tilde{\rm D}_{[b}v_{a]}$ are the kinematic variables of the peculiar motion in the tilded frame. By construction, $\vartheta$ and $\tilde{\vartheta}$ describe the volume expansion/contraction of the drift flow (in the $u_a$ and the $\tilde{u}_a$ frame respectively). Similarly, $\varsigma_{ab}$ and $\tilde{\varsigma}_{ab}$ monitor the associated shear, while $\varpi_{ab}$ and $\tilde{\varpi}_{ab}$ are the two peculiar vorticity tensors.

\begin{figure}[tbp]
\centering \vspace{7cm} \includegraphics{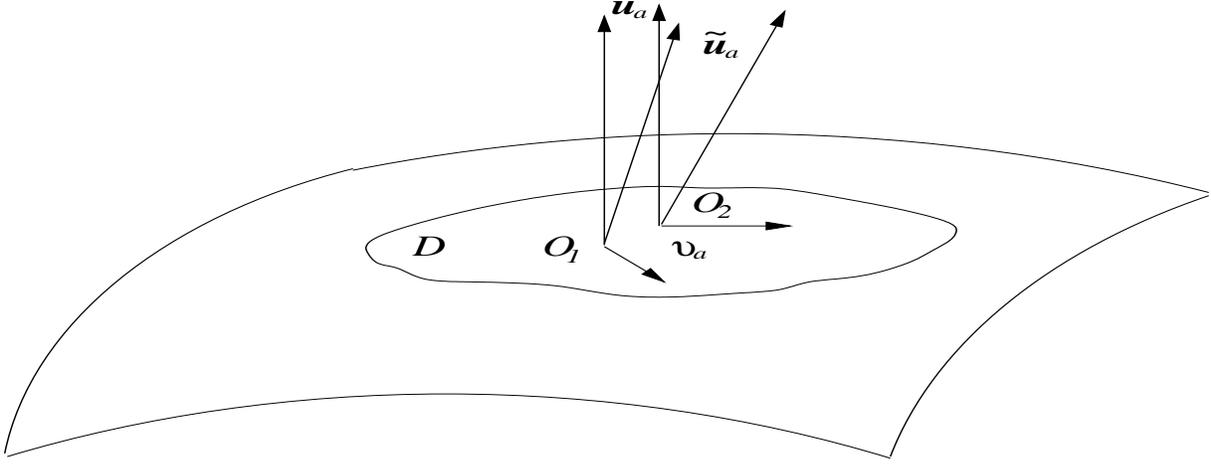} \caption{Neighbouring observers ($O_1$ and $O_2$) inside the bulk flow domain ($D$), with  spacelike 3-velocity $v_a$ tangent to their peculiar flow lines. The 4-velocity field $u_a$ defines the Hubble expansion, while $\tilde{u}_a$ is the 4-velocity of our ``drifting'' observers (see Eq.~(\ref{Lorentz})). The peculiar Raychaudhuri equation (see \S~\ref{sPRE} below) monitors the time evolution of $\vartheta={\rm D}^av_a$ and $\tilde{\vartheta}=\tilde{\rm D}^av_a$, the two scalars measuring the average separation between two neighbouring peculiar flow lines.}  \label{fig:pflow}
\end{figure}

For our purposes, the key variables are $\vartheta$ and $\tilde{\vartheta}$, the scalars monitoring the mean separation between neighbouring peculiar flow-lines (see Fig.~\ref{fig:pflow}), in the reference and the tilded frame respectively. When these are positive, this separation increases and the drift flow is expanding. In the opposite case we have contraction. The deceleration or acceleration of the peculiar expansion/contraction are decided by the time evolution of $\vartheta$ and $\tilde{\vartheta}$, which follows from the Raychaudhuri equation of the drift motion (see \S~\ref{ssRHF} and \S~\ref{ssRDF} below). Finally, we note that, for arbitrarily large peculiar velocities, the aforementioned scalars are related by
\begin{equation}
\tilde{\vartheta}= \vartheta- {1\over3}\,\gamma^2v^2\left(\Theta-\vartheta\right)- \gamma^2\left(v^2A_a-\dot{v}_a\right)v^a- \gamma^2\left(\sigma_{ab}-\varsigma_{ab}\right)v^av^b\,.  \label{thetas}
\end{equation}
When dealing with non-relativistic bulk flows (with $v^2\ll1$), we may only keep up to $v$-order terms on the right-hand side of the above, in which case we conclude that $\tilde{\vartheta}\simeq \vartheta$.

\section{The peculiar Raychaudhuri equation}\label{sPRE}
As mentioned in the beginning, Raychaudhuri's formula is a geometrical relation that can be obtained directly from the Ricci identity. Applying the latter to the 4-velocity of the fundamental observers led to the ``standard'' form of the Raychaudhuri equation (see expressions (\ref{Ray1}) and (\ref{Ray2}) in~\S~\ref{ssSRE}). Here, we will apply the Ricci identities to the 3-velocity of the peculiar motion.

\subsection{Relative to the Hubble frame}\label{ssRHF}
Consider the family of the drifting observers introduced in \S~\ref{ssDOs} earlier. Applying the Ricci identities to the peculiar velocity field gives
\begin{equation}
2\nabla_{[a}\nabla_{b]}v_c= R_{abcd}v^d\,.  \label{vRicci}
\end{equation}
Let us take the view point of the fundamental observers first. Contracting the above along the reference 4-velocity field, taking the trace of the resulting expression and then using decomposition (\ref{Dbva}), provides the Raychaudhuri equation relative to the $u_a$-frame.
\begin{eqnarray}
\dot{\vartheta}&=& -{1\over3}\,\Theta\vartheta- \sigma_{ab}\varsigma^{\,ab}+ \omega_{ab}\varpi^{ab}+ {\rm D}^a\dot{v}_a+ A^a\dot{v}_a- {1\over3}\,\Theta A^av_a- A^av^b\left(\sigma_{ab}-\omega_{ab}\right)\nonumber\\ &{}&-R_{ab}u^av^b\,,  \label{HpRay1}
\end{eqnarray}
where $\dot{v}_a=u^b\nabla_bv_a$. Comparing the above to the ``standard'' expression (\ref{Ray1}), we notice a number of (formalistic) analogies and differences. The first five terms on the right-hand side of (\ref{HpRay1}), in particular, have direct analogues in Eq.~(\ref{Ray1}). The next three terms, on the other hand, are new. Of particular interest is the last term of (\ref{HpRay1}), which carries the explicit effects of spacetime curvature. In contrast to its standard analogue, this term is not contracted twice along the observers time direction, but there is also contraction along the spacelike $v_a$-field. Technically speaking, this ensures that $R_{ab}u^av^b=u^aR_{ab}h^b{}_cv^c= -q_av^a$, where $q_a=-h_a{}^bR_{bc}u^c$ is the energy flux vector of the matter distribution (e.g.~see~\cite{TCM}). Using this last result, Eq.~(\ref{HpRay1}) recasts into
\begin{eqnarray}
\dot{\vartheta}&=& -{1\over3}\,\Theta\vartheta- \sigma_{ab}\varsigma^{\,ab}+ \omega_{ab}\varpi^{ab}+ {\rm D}^a\dot{v}_a+ A^a\dot{v}_a- {1\over3}\,\Theta A^av_a- A^av^b\left(\sigma_{ab}-\omega_{ab}\right)\nonumber\\ &{}&+q_av^a\,.  \label{HpRay2}
\end{eqnarray}
Hence, as seen from the (fundamental/reference) $u_a$-frame, the energy density of the matter component does not explicitly affect the $\vartheta$-evolution. Instead, any direct effects of the matter distribution come from the associated energy flux.

\subsection{Relative to the drifting frame}\label{ssRDF}
Switching to the coordinate system of the drifting observers, we contract Eq.~(\ref{vRicci}) along $\tilde{u}_a$, which defines the corresponding time direction. The trace of the resulting contraction, this time combined with decomposition (\ref{tDbva}), leads to
\begin{eqnarray}
\tilde{\vartheta}^{\prime}&=& -{1\over3}\,\tilde{\Theta}\tilde{\vartheta}- \tilde{\sigma}_{ab}\tilde{\varsigma}^{\,ab}+ \tilde{\omega}_{ab}\tilde{\varpi}^{ab}+ \tilde{\rm D}^av_a^{\prime}+ \tilde{A}^av_a^{\prime}- {1\over3}\,\tilde{\Theta}\tilde{A}^av_a- \tilde{A}^av^b\left(\tilde{\sigma}_{ab}-\tilde{\omega}_{ab}\right)+ \tilde{A}^a\tilde{\rm D}_a\left(\gamma v^2\right)\nonumber\\ &{}&-R_{ab}\tilde{u}^av^b\,,  \label{tpRay1}
\end{eqnarray}
with $v_a^{\prime}=\tilde{u}^b\nabla_bv_a$ (see~\cite{ET} for the details). This is the Raychaudhuri equation of the peculiar flow as seen by the drifting observer. Comparing this relation to expression (\ref{HpRay1}), reveals one new term (the eighth) on the right-hand side. There is also a formalistic difference between the last terms in the two equations. Let us take a closer look at this difference. Noting that $v_a= \tilde{h}_a{}^bv_b-\gamma v^2\tilde{u}_a$, Eq.~(\ref{tpRay1}) becomes
\begin{eqnarray}
\tilde{\vartheta}^{\prime}&=& -{1\over3}\,\tilde{\Theta}\tilde{\vartheta}- \tilde{\sigma}_{ab}\tilde{\varsigma}^{\,ab}+ \tilde{\omega}_{ab}\tilde{\varpi}^{ab}+ \tilde{\rm D}^av_a^{\prime}+ \tilde{A}^av_a^{\prime}- {1\over3}\,\tilde{\Theta}\tilde{A}^av_a- \tilde{A}^av^b\left(\tilde{\sigma}_{ab}-\tilde{\omega}_{ab}\right)+ \tilde{A}^a\tilde{\rm D}_a\left(\gamma v^2\right)\nonumber\\ &{}&- v^a\tilde{h}_a{}^bR_{bc}\tilde{u}^c+ \gamma v^2R_{ab} \tilde{u}^a\tilde{u}^b\,,  \label{tpRay2}
\end{eqnarray}
where now $R_{ab}\tilde{u}^a\tilde{u}^b= (\tilde{\rho}+3\tilde{p})/2-\Lambda$ and $\tilde{h}_a{}^bR_{bc}\tilde{u}^c=-\tilde{q}_a$ respectively represent the total (effective) gravitational energy and the flux, as measured in the tilded frame. Therefore, written in the frame of the drifting observer, the Raychadhuri equation of the peculiar motion reads
\begin{eqnarray}
\tilde{\vartheta}^{\prime}&=& -{1\over3}\,\tilde{\Theta}\tilde{\vartheta}- \tilde{\sigma}_{ab}\tilde{\varsigma}^{\,ab}+ \tilde{\omega}_{ab}\tilde{\varpi}^{ab}+ \tilde{\rm D}^av_a^{\prime}+ \tilde{A}^av_a^{\prime}- {1\over3}\,\tilde{\Theta}\tilde{A}^av_a- \tilde{A}^av^b\left(\tilde{\sigma}_{ab}-\tilde{\omega}_{ab}\right)+ \tilde{A}^a\tilde{\rm D}_a\left(\gamma v^2\right)\nonumber\\ &{}&+\tilde{q}_av^a+ \gamma v^2\left[{1\over2}\,(\tilde{\rho}+3\tilde{p}) -\Lambda\right]\,.  \label{tpRay3}
\end{eqnarray}
The most intriguing effect is perhaps the one encoded in the last term of the above, according to which the roles of the matter and of the cosmological constant are reversed. In particular, as seen form the drifting frame, the presence of conventional matter (with $\tilde{\rho}+3\tilde{p}>0$) tends to accelerate the peculiar expansion, while a positive $\Lambda$ leads to deceleration~\cite{ET}. Although this is an entirely relative motion effect, it marks a potentially significant qualitative difference in the way the peculiar motion appears in the two frames. Nevertheless, the last term in Eq.~(\ref{tpRay3}) becomes important only when the drift velocities are highly relativistic (i.e.~for $v\sim1$).

In cosmology, all the large-scale peculiar velocities reported so far are way below the relativistic limit. Even the dark flows of~\cite{KA-BKE} and the fast bulk motions of~\cite{AF}, which extend to Gpc-order regions, have velocities around hundred times smaller than the speed of light. Such drift motions have $\gamma v^2\sim10^{-4}$, which makes the coefficient of the last term on the right-hand side of (\ref{tpRay3}) very small. The same coefficient drops even further (to values of the order of $10^{-6}$) when applied to the surveys of~\cite{WFH}, which indicate bulk flows around $400$~km/sec on scales close to 100~Mpc. Consequently, for the time being at least, the last term in Eq.~(\ref{tpRay3}) is of theoretical rather than practical interest. Nevertheless, the overall weakness of the matter terms in the peculiar Raychaudhuri equation is a major difference relative to its standard counterpart (compare to expression (\ref{Ray2})), where the matter essentially dictates the average kinematics. As we will see next, the demise of the role of the matter enhances that of the ``peculiar acceleration''.

\section{The case of an almost-FRW universe}\label{sCA-FRWU}
The analysis presented so far has been quite general. Our formulae are fully nonlinear and apply to any spacetime that contains a general (imperfect) fluid and is supplied with arbitrarily large peculiar velocities. In what follows, we will assume non-relativistic drift flows and also linearise these expressions around an FRW cosmology filled with ordinary pressure-free dust.

\subsection{Hubble vis \`a vis drift flow}\label{ssHvavDF}
In relativity, observers moving with respect to each other have a different perception of what we might call reality and the faster the relative motion the greater this difference. Employing the Lorentz boost of Eq.~(\ref{Lorentz}), one can write down the expressions relating all the physical variables measured in the two frames (e.g.~see~\cite{M}). When the relative velocities are small, namely non-relativistic, we have $v^2\ll1$ and $\gamma\simeq1$. In that case, keeping up to $v$-order terms, one can reduce the full equations to a set of considerably simpler relations. Among others, the latter show that the perfect-fluid assumption can only apply to one of the relatively moving frames. In addition, although the 4-acceleration may vanish in one of the aforementioned coordinate systems, it is generally nonzero in the other (see~\cite{M} and also Eqs.~(\ref{lrHDfs1})-(\ref{lrHDfs3}) below).

Most cosmological studies identify their fundamental observers with the CMB frame, which is at rest relative to the smooth Hubble flow. Any other (tilded) observer moves with respect to the aforementioned reference frame and therefore has a different perception of what we may call reality. The Milky Way, for example, is believed to drift at approximately 600~km/sec. Suppose that we live in an almost FRW universe filled with an ideal medium of zero pressure. Then, $p$, $q_a$, $\pi_{ab}$ and $A_a$ are zero to first approximation and observers following the Hubble frame move along timelike geodesics.\footnote{The symmetric and trace-free tensor $\pi_{ab}$ represents the anisotropic fluid pressure. In particular, $\pi_{ab}=h_{\langle a}{}^ch_{b\rangle}{}^dT_{cd}$, where $T_{ab}$ is the energy-momentum tensor of the matter. The isotropic pressure, on the other hand, is given by $p=h^{ab}T_{ab}/3$. Also, $q_a= -h_a{}^bT_{bc}u^c$ and $\rho=T_{ab}u^au^b$ (e.g.~see~\cite{TCM}).} As a result, the linear transformation laws between the Hubble and the tilded frame of a drifting observer reduce to~\cite{M}
\begin{eqnarray}
&\tilde{\rho}= \rho\,, \hspace{5mm} \tilde{p}= 0\,, \hspace{5mm} \tilde{q}_a= -\rho v_a\,, \hspace{5mm} \tilde{\pi}_{ab}= 0\,,& \label{lrHDfs1}\\ &\tilde{\Theta}= \Theta+ \tilde{\vartheta}\,, \hspace{5mm} \tilde{\omega}_{ab}= \omega_{ab}+ \tilde{\varpi}_{ab}\,,&  \label{lrHDfs2}\\ &\tilde{\sigma}_{ab}= \sigma_{ab}+ \tilde{\varsigma}_{ab}\,, \hspace{5mm} \tilde{A}_a= v^{\prime}_a+ H v_a\,.&  \label{lrHDfs3}
\end{eqnarray}
As mentioned before, although the perturbed fluid remains ideal and pressure-free, there is a nonzero effective heat flux and a non-vanishing 4-acceleration in the $\tilde{u}_a$-frame. Both vectors are the direct result of the observers' relative motion.\footnote{One could in principle switch off the heat flux and the 4-acceleration in the drifting frame (i.e.~set $\tilde{q}_a=0=\tilde{A}_a$. In that case, $q_a$ and $A_a$ will generally take nonzero values and the relative motion effects will appear in the Hubble frame. However, the latter is the fundamental reference system in which the dust is at rest. It therefore makes more physical sense to set $q_a$ and $A_a$ to zero, instead of their tilded counterparts.} Drift flows also affect the average volume expansion, induce vorticity and trigger additional shear distortions (see Eqs.~(\ref{lrHDfs2}) and (\ref{lrHDfs3}a) above, as well as~\cite{CT} for a sample of related studies and additional references).

\subsection{Following the drift flow}\label{ssFDF}
Consider a perturbed Friedmann universe filled with pressureless dust and allow for a family of realistic observers, moving with non-relativistic peculiar velocity ($v_a$) relative to the mean (Hubble) expansion. In the unperturbed background peculiar velocities vanish, which makes $v_a$ a gauge-invariant, first-order variable. Also, the high symmetry of the FRW spacetime is only compatible with scalars that depend solely on time. Thus, when dealing with dust, the only zero-order variables are the Hubble parameter ($H$), the matter density ($\rho$) and the Ricci scalar ($\mathcal{R}$) of the 3-dimensional hypersurfaces. All the rest will be treated as linear perturbations and terms of perturbative order higher than the first will be dropped. Applying these principles to Eq.~(\ref{tpRay3}), the latter reduces to the first-order expression
\begin{equation}
\tilde{\vartheta}^{\prime}= -H\tilde{\vartheta}+ \tilde{\rm D}^av_a^{\prime}\,.  \label{ltpRay1}
\end{equation}
The above is the linear Raychaudhuri equation of a drift flow that moves relative to the universal expansion with peculiar velocity $v_a$. Expression (\ref{ltpRay1}) monitors the sign and the magnitude of $\tilde{\vartheta}^{\prime}$, which together with those of $\tilde{\vartheta}$ determine the mean peculiar kinematics. As mentioned earlier (see Eq.~(\ref{tDbva}) in \S~\ref{ssDF}), the peculiar motion is expanding (on average) when $\tilde{\vartheta}$ is positive. In the opposite case, we are dealing with a contracting drift flow. Therefore, positive values for $\tilde{\vartheta}^{\prime}$ imply an accelerating peculiar expansion (when $\tilde{\vartheta}$ is also positive), or a decelerating peculiar contraction (when $\tilde{\vartheta}$ is negative). For negative values of $\tilde{\vartheta}^{\prime}$ the situation is reversed. When $\tilde{\vartheta}$ is negative, we have drift flows that contract with an ever increasing speed, while positive values of $\tilde{\vartheta}$ indicate bulk motions that expand at a decelerating pace.

The absence of explicit matter terms from Eq.~(\ref{ltpRay1}), means that (to linear order) the average peculiar kinematics is not directly affected by the local gravitational field. Also, given that $H>0$ at all times, the first term on the right-hand side of (\ref{ltpRay1}) always plays the role of friction, by slowing down either the expansion (when $\tilde{\vartheta}>0$) or the contraction (when $\tilde{\vartheta}<0$) of the drift flow. The effect of the last term in (\ref{ltpRay1}) is less straightforward to decode and estimate. Currently, there are peculiar-velocity measurements on large scales, though not in mutual agreement~\cite{WFH}-\cite{AF}. From these data, one can derive estimates for the possible values of $\tilde{\vartheta}$ on certain scales (see \S~\ref{ssSNEs} below). So far, however, there are no reports on peculiar accelerations. In other words, we have observational estimates for $v_a$ but not for $v_a^{\prime}$. This makes the last term on the right-hand side of Eq.~(\ref{ltpRay1}) highly ambiguous, in terms of magnitude and impact. Next, we will attempt to extract some additional information by turning to cosmological perturbation theory.

\subsection{The role of inhomogeneity}\label{ssRI}
Peculiar velocities are believed to be the result of structure formation and of the increasing inhomogeneity and anisotropy of the universe. More specifically, drift flows are related to density inhomogeneities. In the drifting frame, the latter are monitored by the orthogonally projected gradient $\tilde{\Delta}_a= (a/\rho)\tilde{{\rm D}}_a\tilde{\rho}$, which describes density variations between two neighbouring observers~\cite{TCM}. To obtain the linear relation between peculiar velocities and density inhomogeneities, we first recall that in the drifting frame the cosmic fluid is no longer perfect. Following Eqs.~(\ref{lrHDfs1}), the tilded observer still sees a pressure-free medium. However, the latter has now a nonzero (effective) energy-flux vector that is entirely a relative motion effect. All these mean that, in the frame of the drift flow, density inhomogeneities are described by the linear expression~\cite{TCM}
\begin{equation}
\tilde{\Delta}^{\prime}_a= -\tilde{\mathcal{Z}}_a+ {3aH\over\rho}\left(\tilde{q}^{\,\prime}_a+4H\tilde{q}_a\right)- {a\over\rho}\,\tilde{\rm D}_a\tilde{\rm D}^b\tilde{q}_b\,,  \label{ltpDelta1}
\end{equation}
where $\tilde{\mathcal{Z}}_a=a\tilde{\rm D}_a\tilde{\Theta}$ describes spatial variations in the average expansion of the universe.\footnote{In the Hubble frame the last two terms on the right-hand side of Eq.~(\ref{ltpDelta1}) are identically zero. Also, $\tilde{\Delta}_a$ should be replaced by $\Delta_a=(a/\rho){\rm D}_a\rho$, $\mathcal{Z}_a$ by $\mathcal{Z}_a=a{\rm D}_a\Theta$ and $\tilde{\Delta}_a^{\prime}$ by $\dot{\Delta}_a= u^b\nabla_b\Delta_a$. Therefore, the standard linear solutions of cosmological perturbation theory do not generally hold in the tilded frame.} Note that $\tilde{q}_a=-\rho v_a$ to linear order (see (\ref{lrHDfs1}c) in \S~\ref{ssHvavDF}), which ensures that $\tilde{q}^{\,\prime}_a=3H\rho v_a-\rho v^{\prime}_a$ at the same perturbative level (recall that $\rho^{\,\prime}=\dot{\rho}= -3H\rho$ in the background). Using these auxiliary relations and keeping in mind that $\tilde{\vartheta}=\tilde{\rm D}^av_a$, the (comoving) 3-divergence of (\ref{ltpDelta1}) leads to
\begin{equation}
\tilde{\Delta}^{\,\prime}= -\tilde{\mathcal{Z}}- 3a^2H \left(\tilde{\rm D}^av_a^{\prime}+H\tilde{\vartheta}\right)+ a^2\tilde{\rm D}^2\tilde{\vartheta}\,,  \label{ltDelta2}
\end{equation}
with $\tilde{\Delta}=a\tilde{\rm D}^a\tilde{\Delta}_a$ by definition and given that $\tilde{\Delta}^{\prime}=a\tilde{\rm D}^a \tilde{\Delta}^{\prime}_a$ to first order~\cite{TCM}. Similarly, $\tilde{\mathcal{Z}}=a\tilde{\rm D}^a\tilde{\mathcal{Z}}_a$ and $\tilde{\rm D}^2=\tilde{\rm D}^a\tilde{\rm D}_a$ is the Laplacian operator in the 3-D space of the drifting observer.\footnote{The gradient $\tilde{\Delta}_a$ contains collective information for all three types of density inhomogeneities, namely scalar, vector and tensor. The scalar part, in particular, is monitored by the comoving divergence $\tilde{\Delta}=a\tilde{\rm D}^a\tilde{\Delta}_a$ and corresponds to what we normally call density perturbations, that is overdensities or underdensities in the matter distribution (e.g.~see~\cite{TCM}). An exactly analogous description also applies to $\tilde{\mathcal{Z}}_a$ and $\tilde{\mathcal{Z}}$.}

The expansion gradient ($\tilde{\mathcal{Z}}_a$) satisfies the so-called shear constraint (see Eq.~(1.3.6) in \S~1.3 of~\cite{TCM}), which in the drifting frame linearises to
\begin{equation}
2\tilde\mathcal{Z}_a=3a\left(\tilde{\rm D}^a\tilde{\sigma}_{ab} -{\rm curl}\tilde{\omega}_a-\rho v_a\right)\,.  \label{ltshearcon1}
\end{equation}
Here, ${\rm curl}\tilde{\omega}_a=\tilde{\varepsilon}_{abc} \tilde{\rm D}^b\tilde{\omega}^c$ by definition, $\tilde{\varepsilon}_{abc}=\eta_{abcd}\tilde{u}^d$ is the (totally antisymmetric) 3-D Levi-Civita tensor and $\eta_{abcd}$ represents its 4-D counterpart. Taking the comoving 3-divergence of the above and keeping up to linear order terms, we arrive at
\begin{equation}
2\tilde{\mathcal{Z}}= 3\tilde{\Sigma}- 9a^2H^2\Omega\tilde{\vartheta}\,,  \label{ltshearcon2}
\end{equation}
since $\tilde{\rm D}^a{\rm curl} \tilde{\omega}_a=0$ to first approximation. In deriving the above, we have also used the background relation $\Omega=\rho/3H^2$ for the density parameter and introduced the scalar $\tilde{\Sigma}=a^2\tilde{\rm D}^a \tilde{\rm D}^b\tilde{\sigma}_{ab}$. The latter provides a measure of the spatial anisotropy of our model. Combining (\ref{ltDelta2}) and (\ref{ltshearcon2}), one can eliminate $\tilde{\mathcal{Z}}$ from the system and thus recast expression (\ref{ltDelta2}) into
\begin{equation}
\tilde{\Delta}^{\prime}= -{3\over2}\,\tilde{\Sigma}- 3a^2H\left[\tilde{\rm D}^av^{\prime}_a +H\left(1-{3\over2}\,\Omega\right)\tilde{\vartheta}\right]+ a^2\tilde{\rm D}^2\tilde{\vartheta}\,.  \label{ltpDelta3}
\end{equation}
The above describes the evolution of linear scalar perturbations in the matter distribution, namely overdensities or underdensities, relative to the drifting (the tilded) frame. Solving this expression for $\tilde{\rm D}^av^{\prime}_a$ and then substituting the result into the right-hand side of Eq.~(\ref{ltpRay1}), gives
\begin{equation}
\tilde{\vartheta}^{\prime}= -2H\left(1-{3\over4}\,\Omega\right)\tilde{\vartheta}+ {1\over3H}\,\tilde{\rm D}^2\tilde{\vartheta}- {1\over3a^2H}\left(\tilde{\Delta}^{\prime} +{3\over2}\,\tilde{\Sigma}\right)\,.  \label{ltpRay2}
\end{equation}
This is a new version of the ``peculiar'' Raychaudhuri equation (compare to expression (\ref{ltpRay1}) in \S~\ref{ssFDF}), which now explicitly incorporates the effects of spatial inhomogeneity and anisotropy. We also note the appearance of the background density parameter ($\Omega$) and the Laplacian term on the right-hand side of the above. The latter implies that the value of $\tilde{\vartheta}^{\prime}$, namely the deceleration/acceleration rate of the peculiar motion, is scale dependent.

\subsection{The scale-dependent formula}\label{ssSDF}
We may decode the scale dependence of Eq.~(\ref{ltpRay2}) by harmonically decomposing all the first-order variables. Introducing the scalar harmonics $\tilde{\mathcal{Q}}^{(n)}$, where $\tilde{\mathcal{Q}}^{\prime\,(n)}=0$ and $\tilde{\rm D}^2 \tilde{\mathcal{Q}}^{(n)}=-(n/a)^2\tilde{\mathcal{Q}}^{(n)}$, we may write $\tilde{\vartheta}=\sum_n\tilde{\vartheta}_{(n)} \tilde{\mathcal{Q}}^{(n)}$, $\tilde{\Delta}= \sum_n\tilde{\Delta}_{(n)}\tilde{\mathcal{Q}}^{(n)}$ and $\tilde{\Sigma}=\sum_n\tilde{\Sigma}_{(n)} \tilde{\mathcal{Q}}^{(n)}$, with $\tilde{\rm D}_a\tilde{\vartheta}_{(n)}=0=\tilde{\rm D}_a \tilde{\Delta}_{(n)}=\tilde{\rm D}_a\tilde{\Sigma}_{(n)}$. Then, expression (\ref{ltpRay2}) becomes
\begin{equation}
\tilde{\vartheta}_{(n)}^{\prime}= -2H^2\left\{\left[1-{3\over4}\,\Omega+{1\over6}\, \left({\lambda_H\over\lambda_n}\right)^2\right] {\tilde{\vartheta}_{(n)}\over H}+ {1\over6}\left({\lambda_H\over\lambda_K}\right)^2 \left({\tilde{\Delta}_{(n)}^{\prime}\over H}+ {3\over2}\,{\tilde{\Sigma}_{(n)}\over H}\right)\right\}\,.  \label{ltpRay3}
\end{equation}
where the quantity in braces is now dimensionless. Note that $\lambda_H=1/H$ is the Hubble length, $\lambda_n=a/n$ is the perturbed scale (with $n$ representing the comoving wavenumber of the perturbed mode) and $\lambda_K=a$ is the curvature radius of the universe. The above is the main result of our analysis, providing the linearised Raychaudhuri equation of a peculiar flow in a perturbed FRW cosmology that contains nothing else but pressure-free dust. Not surprisingly, expression (\ref{ltpRay3}) shows that the rate of the average deceleration/acceleration of the drift flow is scale dependent and affected by the spatial inhomogeneity and anisotropy of the adopted cosmological model.

Looking at the right-hand side of Eq.~(\ref{ltpRay3}), the effect of the friction term (the first inside the braces) increases on subhorizon scales, where $\lambda_H/\lambda_n\gg1$. In fact, on sufficiently small scales the coefficient of this term is entirely dominated by the scale ratio $\lambda_H/\lambda_n$. Having said that, we should also point out that on very small scales (with $\lambda_n\ll100$~Mpc) our linear analysis is expected to break down. There is also an $\Omega$-dependence in (\ref{ltpRay3}), which can in principle change the sign of the friction term. Nevertheless, the background density parameter makes a real difference only on supper-Hubble lengths (where $\lambda_H/\lambda_n\ll1$) and in closed FRW models with $\Omega>4/3$. Note that the recently released Planck data suggest that $|1-\Omega|\lesssim10^{-3}$~\cite{Aetal2}. Therefore, the friction term on the right-hand side of (\ref{ltpRay3}) generally slows down the average expansion/contraction of a bulk peculiar flow and the decelerating effect increases with decreasing scale.

Estimating the strength of the inhomogeneity/anisotropy term (the second inside the braces of Eq.~(\ref{ltpRay3})) is a more complicated process, which goes beyond the scope of this work. It should be noted though that $(\lambda_H/\lambda_K)^2= |1-\Omega|/|K|$, where $K=\pm1$ is the 3-curvature index of the FRW background. Current observations indicate that $|1-\Omega|\lesssim10^{-3}$, which implies that $(\lambda_H/\lambda_K)^2\lesssim10^{-3}$ as well. It also sounds plausible to argue that $|\tilde{\Sigma}_{(n)}/H|\ll1$ throughout the linear regime. On these grounds, it sounds plausible to argue that the second term inside the braces of Eq.~(\ref{ltpRay3}) has a negligible overall effect. We should also keep in mind, however, that (to the best of our knowledge) there are no observational (or theoretical) constrains for the ratio $\tilde{\Delta}_{(n)}^{\prime}/H$, which could in principle take relatively large (absolute) values.

\subsection{Solutions and numerical estimates}\label{ssSNEs}
Equation (\ref{ltpRay3}) can be solved analytically in the simplest case, when the effects of inhomogeneity and anisotropy are negligible. Then, the peculiar Raychaudhuri equation reduces to
\begin{equation}
\tilde{\vartheta}^{\prime}\simeq -2H\left[1-{3\over4}\,\Omega +{1\over6}\,\left({\lambda_H\over\lambda}\right)^2\right] \tilde{\vartheta}\,,  \label{ltpRay4}
\end{equation}
where the mode-index ($n$) has been dropped for ``economy'' reasons. When $\Omega\leq4/3$, the above ensures a decelerating peculiar expansion/contraction, when $\vartheta$ is positive/negative respectively (i.e.~it gives $\tilde{\vartheta}^{\prime}<0$ when $\tilde{\vartheta}>0$ and $\tilde{\vartheta}^{\prime}>0$ for $\tilde{\vartheta}<0$). Also, the smaller the scale of the drift motion, the stronger the decelerating effect of the Hubble flow. Finally, ignoring the time-dependence of the coefficient on the right-hand side of (\ref{ltpRay4}), which is a fair approximation for a brief period in the (late) expansion of the universe, leads to \begin{equation}
\tilde{\vartheta}= \tilde{\vartheta}_0 \left({a_0\over a}\right)^{2\beta}\,,  \label{tvartheta1}
\end{equation}
with $\beta=[1-3\Omega/4+(\lambda_H/\lambda)^2/6]$ and the zero suffix indicating a given initial time. Thus, on scales close to Hubble horizon and on nearly flat FRW backgrounds (i.e.~for $\lambda\simeq\lambda_H$ and $\Omega\simeq1$), the volume scalar of the peculiar flow decays as $\tilde{\vartheta}\propto a^{-5/6}\propto t^{-5/9}$ (recall that $a\propto t^{2/3}$ during the dust era). Well inside the Hubble radius, on the other hand, the decay is much faster. Put another way, the background expansion slows down the bulk motion, irrespective of whether the latter expands or contracts on average, and the decelerating effect is stronger on small scales.

Let us go back to Eq.~(\ref{ltpRay4}). From the observational point of view, it is not the value of $|\tilde{\vartheta}^{\prime}|$ that is important but that of the dimensionless ratio $|\tilde{\vartheta}^{\prime}/\dot{H}|$.~\footnote{Following Eq.~(\ref{ltpRay4}), $\tilde{\vartheta}$ and $\tilde{\vartheta}^{\prime}$ have opposite signs, since $H>0$ always and $\Omega\simeq1$. Then, in accord with (\ref{ltpRay5}) and given that $\dot{H}<0$ always, $\tilde{\vartheta}^{\prime}/ \dot{H}<0$ when the bulk flow is contracting (i.e.~for $\tilde{\vartheta}<0$) and $\tilde{\vartheta}/\dot{H}>0$ otherwise.} The latter provides a measure of the mean deceleration of the bulk flow, relative to that of the background universe. Therefore, dividing (\ref{ltpRay4}) by $\dot{H}$, we arrive at
\begin{equation}
\left|{\tilde{\vartheta}^{\prime}\over\dot{H}}\right|\simeq 2\left(1+{1\over2}\,\Omega\right)^{-1} \left[1-{3\over4}\,\Omega +{1\over6}\,\left({\lambda_H\over\lambda}\right)^2\right] \left|{\tilde{\vartheta}\over H}\right|\,,  \label{ltpRay5}
\end{equation}
since $\dot{H}=-H^2(1+\Omega/2)$ to zero order for dust~\cite{TCM}. Given the value of $|\tilde{\vartheta}/H|$ on a certain scale, one could use the above to estimate the ratio $|\tilde{\vartheta}^{\prime}/\dot{H}|$ and thus the relative deceleration of the bulk flow in question. For instance, the surveys of~\cite{WFH} report peculiar velocities around 400~km/sec on scales close to 100/$h$~Mpc. Since $\tilde{\vartheta}={\rm D}^av_a$ by definition and given that $\Omega\simeq1$, we can replace the covariant derivatives with partials and use the approximate relation $|\tilde{\vartheta}|\simeq |\partial^av_a|\simeq3v/r$. Note that $v$ is the magnitude of the reported peculiar velocity and $r$ is the size the bulk flow in question (with $r\simeq\lambda$). Then, $|\tilde{\vartheta}/H|\simeq 3v/rH$, which translates into $|\tilde{\vartheta}/H|\simeq1/10$ for the data of~\cite{WFH}. The latter correspond to a scale ratio of $\lambda_H/\lambda\simeq20$. Substituting these numbers back into (\ref{ltpRay5}), we find that $|\tilde{\vartheta}^{\prime}/ \dot{H}|\simeq10$ for a nearly flat universe.~\footnote{We remind the reader that these numerical estimates have been obtained after dropping the second term inside the braces of Eq.~(\ref{ltpRay3}). In other words, we have assumed that on the wavelengths in question the non-uniformity of the universe has a negligible effect. It is conceivable, however, that on scales close to 100~Mpc this may not be the case. The ratio $|\tilde{\Delta}^{\prime}/H|$, in particular, may be large enough to make a difference.} Repeating this process for the data of~\cite{KA-BKE,AF}, reporting peculiar velocities around $1000$~km/sec on Gpc-order scales (i.e.~setting $|\tilde{\vartheta}/H|\simeq3/70$ and $\lambda_H/\lambda\simeq3$ in Eq.~(\ref{ltpRay5})), gives $|\tilde{\vartheta}^{\prime}/ \dot{H}|\simeq1/10$.

Overall, the scale dependence of (\ref{ltpRay5}) means that on subhorizon lengths the ratio $\tilde{\vartheta}^{\prime}/\dot{H}$, as measured by an observers living inside the bulk flow, is larger (in absolute values) than $\tilde{\vartheta}/H$, typically by one or two orders of magnitude. When combined with the data, this means that on scales of few hundred Mpc the deceleration rate of weakly expanding/contracting bulk motions (i.e.~drift flows with $|\tilde{\vartheta}/H|<1$) can match or even surpass that of the background expansion. This is not entirely surprising, because small values for $|\tilde{\vartheta}/H|$ do not necessarily guarantee the same for the ratio of the derivatives. Nevertheless, when $|\tilde{\vartheta}^{\prime}/\dot{H}|$ becomes of order unity, an unsuspecting drifting observer might be misled into interpreting certain kinematic observations inaccurately. It would be of particular interest, for example, to examine whether a relatively large value of $|\tilde{\vartheta}^{\prime}/\dot{H}|$ can affect the deceleration parameter measured by observers moving with respect to the smooth Hubble flow (e.g.~see~\cite{T} for a discussion). Investigating the likelihood and the implications of such potential misinterpretations will be the subject of our future work.

\section{Discussion}\label{sD}
Large scale peculiar motions are believed to be the result of structure formation and a direct sign of the increasing spatial inhomogeneity and anisotropy of the post-recombination universe. No realistic observer follows the Hubble expansion, but all move relative to it. Our Milky Way and the Local Group, in particular, drift at a speed of roughly 600~km/sec. Similar, or even faster, peculiar velocities and on considerably larger scales have been recently reported by a number of research groups. Although the surveys disagree on the magnitude of the velocities, they seem to agree on the direction of these large-scale bulk flows. It is well known that relatively moving observers have different perception of reality and the faster the relative motion the greater this difference. Therefore, in principle at least, observers living in a typical galaxy, like the Milky Way for example, see a different universe than those following the smooth Hubble flow. It is therefore of theoretical and practical interest to study the kinematics/dynamics of peculiar motions and investigate their potential effects on the interpretation of the observational data.

We have taken a step in this direction, by considering the mean kinematics of large-scale bulk motions and writing down the associated Raychaudhuri equation. The latter describes changes in the average separation between two neighbouring peculiar-flow lines and in particular whether the bulk flow expands/contracts at a decelerating or an accelerating pace. We have expressed the peculiar Raychaudhuri equation with respect to two frames: one following the Hubble expansion and another drifting along with the bulk flow in question. Without assuming any spacetime symmetries, allowing for arbitrarily large peculiar velocities and considering a general imperfect fluid, we have written down the fully nonlinear expressions of Raychaudhuri's formula in both of the aforementioned coordinate systems. It appears that the most important qualitative difference between the peculiar and the standard forms of the Raychaudhuri equation is in the effects of the matter. The familiar terms, which express the decelerating effect of conventional matter and the accelerating role of a positive cosmological constant, do not appear in the peculiar Raychaudhuri equation or are drastically modified. In the fundamental (reference) frame the only direct effects of the matter component come through its energy-flux vector. As seen from drifting frame, on the other hand, the roles of the ordinary matter and of the cosmological constant are reversed. The energy density and pressure of conventional matter seems to accelerate the peculiar expansion, while a positive cosmological constant appears to do the opposite. Nevertheless, al these relative motion effects are negligible, unless the peculiar velocities are highly relativistic.

Assuming non-relativistic bulk motions and an FRW background universe, filled with ordinary pressure-free dust, we linearised our nonlinear expressions. We did so in the drifting frame, since this is the coordinate system of any realistic observer. When linearing, it is important to recognise that in our frame of reference the cosmic fluid is no longer perfect and that the observers' worldlines are not geodesics (even in the absence of fluid pressure). Initially, this extra complication adds only one term to the right-hand side of the peculiar Raychaudhuri equation, which depends on the peculiar acceleration (the proper-time derivative of the peculiar velocity). The latter is related to the spatial inhomogeneity and anisotropy of our perturbed model. Hence, employing linear cosmological perturbation theory, we arrived at a scale-dependent version of the peculiar Raychaudhuri equation, which also reflects the non-uniformity of our universe.

Broadly speaking, the mean linear kinematics of large-scale bulk flows are determined by the background expansion and by the spatial inhomogeneity/anisotropy of the perturbed (almost-FRW) cosmos. Given that the expansion always acts as an effective source of ``friction'', its role is fairly straightforward to estimate (both qualitatively and quantitatively). To be precise, the Hubble flow slows down the expansion/contraction of the peculiar motion and its decelerating effect increases with decreasing scale. The presence of inhomogeneities and anisotropies, on the other hand, adds extra degrees of freedom that are less straightforward to decode. Nevertheless, on scales larger than 100~Mpc, effects coming from the overall non-uniformity of the universe are probably secondary to those due to the background expansion. Without the inhomogenous/anistropic terms, the peculiar Raychaudhuri equation can be solved analytically and the decelerating role of the Hubble expansion can be quantified in terms of scale. Using data from recent peculiar-velocity surveys, one can also estimate the ratio $|\tilde{\vartheta}^{\prime}/\dot{H}|$. The latter provides the average deceleration of the associated bulk motions relative to that of the background universe, as measured in the ``tilded'' frame. Our results indicate that on scales of few hundred Mpc, the aforementioned ratio can be of order unity, a value large enough to interfere with the way a drifting observer may interpret certain cosmological data. It is our aim to investigate this possibility further in the very near future.

\end{document}